\documentclass[%
 reprint,
 amsmath,amssymb,
 aps,
]{revtex4-2}

\usepackage{graphicx}
\usepackage{dcolumn}
\usepackage{bm}

\begin{document}

\preprint{APS/123-QED}

\title{Radiation of Optical Angular Momentum from a Dipole Source in a Magneto-birefringent disordered Environment}


\author{R. Le Fournis}
\email[]{romuald.lefournis@lpmmc.cnrs.fr}

\author{B.A. van Tiggelen}
\email[]{Bart.Van-Tiggelen@lpmmc.cnrs.fr}
\affiliation{Univ. Grenoble Alpes, CNRS, LPMMC, 38000 Grenoble, France}




\date{\today}

\begin{abstract}
\textbf{Abstract}: We investigate the radiation of optical angular momentum by a dipole gas under uniform magnetic field with an unpolarized source at its center. Conservation of angular momentum implies a torque on both the source and the surrounding environment. We study the separate spin and orbital contributions to the radiated angular momentum. 
 
\end{abstract}

\maketitle


\section{Introduction}

The emission of radiation from a dipole source, such as an atom or a quantum emitter, can be significantly affected by its environment. An important phenomenon in this context is the Purcell effect \cite{Purcell}. It refers to the modification of the radiative properties of a source when placed in close proximity to an environment with specific optical properties. Notably, it has been shown theoretically and experimentally that the angular momentum emitted by a dipole source is affected by its environment \cite{vanTiggelen2020Sep}. Previous work has shown that a dipole source at the center of a Mie sphere under a magnetic field can experience a torque that results in the emission of angular momentum into space \cite{vanTiggelen2023Jan}. In this work we calculate the angular momentum emitted by an unpolarized source placed at the center of a magneto-birefringent heterogeneous environment composed of small electric dipole scatterers. Due to the conservation of angular momentum, not only the source experiences a torque but also the environment. We study the decomposition of the radiated angular momentum into spin and orbital components. Our aim is to understand the influence of multiple scattering of light on both the radiated angular momentum and the torque on the medium.

\section{Radiation of optical angular momentum}

We consider an unpolarized dipole source in the center of a spherical environment with radius $R$ containing a gas of $N$ point-like electric dipoles (Fig.\ref{picture}). The electric dipoles have polarizability \cite{Barron2004Sep}:
 \begin{equation}
     \bm{\alpha}(\omega,\bm{B}_0) = \frac{\alpha(0)\omega^2_0}{\omega^2_0 - (\omega + \omega_c i\bm{\epsilon} \cdot \hat{\bm{B}}_0)^2 - i\gamma \omega}
     \label{PolarizabilityB}
 \end{equation}
where $\gamma$ is the radiative damping rate of the dipoles, $\omega_0$ is the resonance frequency and $\omega_c = eB_0/2m c_0$ is the cyclotron frequency; $\alpha(0)$ is the static polarizability and can be related to the volume $u$ of the dipoles as $\alpha(0)=3u$ \cite{vanTiggelen2021May}. In a microscopic theory for a two-level atom, $\hbar\omega_c$ represents the Zeeman splitting of the excited states due to the interaction of the atom with the magnetic field. The continuity equation for the angular momentum $\bm{J}(r,t)$ confined in a sphere of radius $r$ at time $t$ around the source is given by \cite{Landau1980Jan}:
\begin{figure}[]
  \centering
  \includegraphics[scale = 0.2]{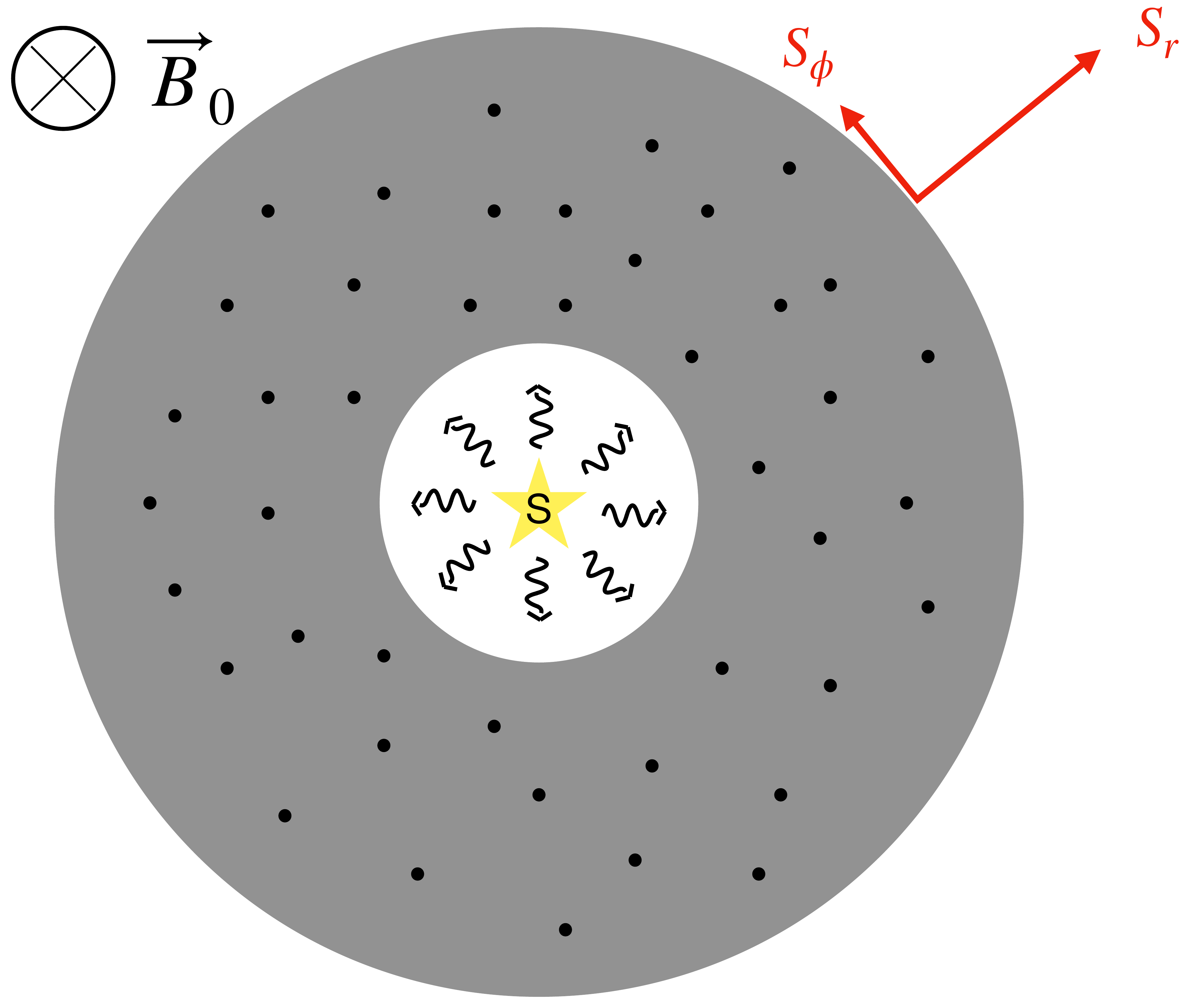}
  \caption{Schematical picture of the geometry considered in this work. An unpolarized dipole source is located at the origin, surrounded by two shells: the first shell is empty to avoid scattering close to the source, while the second shell contains an on average homogeneous dipole gas. The external magnetic field is directed perpendicular to the plane. The electromagnetic Poynting vector exhibits an azymuthal component $\bm{S}_\phi$ which is orthogonal to both the magnetic field and the radial vector. This component induces the emission of angular momentum into space.}
  \label{picture}
\end{figure}
\begin{align}
    \frac{d}{dt} \biggl[ J_k^{\text{mech}} + J_k^{\text{rad}} \biggr](r,t) &=\frac{r^3}{8\pi}\epsilon_{klm}   \nonumber \\
     \times \text{Re}  &\int d \hat{\bm{r}}' \, \hat{r}'_j \hat{r}'_l (E_j \bar{E}_m  + B_j \bar{B}_m)(r \hat{\bm{r}}',t) \nonumber \\
    & \equiv M_{L,k}
    \label{balanceequation}
\end{align}
The right-hand side of this equation represents the optical angular momentum (AM) radiated into space. For $r>R$ the mechanical AM is independent on distance $r$, and $d \bm{J}^{\text{mech}}/dt$ gives the torque on both source and environment. If the source has been stationary for a time much longer than it takes light to travel a distance equal to the radius $r$ ($t \gg r/c_0$), the total electromagnetic AM inside the sphere of radius $r$ stays constant, so that $d \bm{J}^{\text{rad}} (r,t)/dt = \bm{0}$. In that case, the leakage is independent on the distance $r$. According to Eq.~\eqref{balanceequation}, matter experiences a torque equal to the emitted AM. To determine the amount of AM emitted, we can evaluate the right-hand side of Eq.~\eqref{balanceequation} at a large distance from the environment ($r \rightarrow \infty$). The electric and magnetic fields in Eq.~\eqref{balanceequation} are determined by solving the Helmholtz equation for a classical monochromatic source with frequency $\omega = k c_0$ \cite{Newton2013Jun}. The solutions for the electric and magnetic fields can be expressed as,

\begin{equation*}
    \bm{E}(\bm{r}) =  \bm{G}^E(\bm{r})  \cdot (-4\pi i \omega/c^2_0)\bm{j}_S
\end{equation*}
\begin{equation}
    \bm{B}(\bm{r}) =  \bm{G}^B(\bm{r})  \cdot (-4\pi i \omega/c^2_0)\bm{j}_S
    \label{electricfieldmagneticfield}
\end{equation}
We define the Green's functions as $\bm{G}^E(\bm{R})$ and $\bm{G}^B(\bm{R})$ according to
\begin{equation}
    \bm{G}^E(\bm{r}) = \bm{G}_{0}(\bm{r}) + \sum_{\alpha, \beta} \bm{G}_{0}(\bm{r} - \bm{R}_\alpha) \cdot \bm{T}_{\alpha \beta} \cdot \bm{G}_{0}(\bm{R}_\beta)
\end{equation}
\begin{equation}
    \bm{G}^B(\bm{r}) = \bm{G}^0_B(\bm{r}) + \sum_{\alpha,\beta} \bm{G}^0_B(\bm{r} - \bm{R}_\alpha) \cdot \bm{T}_{\alpha \beta} \cdot \bm{G}_0(\bm{R}_\beta)
\end{equation}
To distinguish between the spatial components and the dipole indices of the tensors, we use Latin and Greek indices, respectively. The dot denotes the contraction of Latin indices associated with polarization. The Green's functions $\bm{G}^{0}$ and $\bm{G}^0_{B}$ describe how the electromagnetic waves generated by the electric dipole source propagate in free space and can be obtained from \cite{Landau1980Jan}. For a monochromatic electric dipole source at the origin with a dipole moment $\bm{d}$, the source current is $\bm{j}_S(\bm{r}) = -i \omega \bm{d} \delta(\bm{r})$. The transmission matrix $\bm{T}_{\alpha \beta}$, that appears in both Green's functions, is a $3N \times 3N$ matrix. It describes how light emanating from a source and interacting with a gas of electric dipoles propagates through the environment. The factor 3 corresponds to the 3 polarization states of light. The T-matrix can be calculated as described in \cite{vanTiggelen2021May}. We will first put $\bm{B}_0 = \bm{0}$ and treat the external magnetic field later in perturbation. Without external magnetic field, the T-matrix is given by:
\begin{equation}
    \lbrace \bm{T}^{(0)}_{\alpha \beta} \rbrace =  \frac{t_0(\omega)}{\lbrace \bm{1}\delta_{\alpha \beta} - t_0(\omega)\bm{G}^0(\bm{R}_\alpha - \bm{R}_\beta )(1 - \delta_{\alpha \beta})\rbrace} 
\end{equation}
The superscript $(0)$ indicates that this is the T-matrix in the absence of an external magnetic field; $t_0(\omega)$ is the T-matrix of a single electric dipole and is expressed near its resonance frequency $\omega_0 = c_0 k_0$ as:
\begin{equation}
    \bm{t}_0(\omega) = - \bm{\alpha}(\omega,\bm{B}_0 = \bm{0}) k^2 =  \frac{3 \pi}{k_0}\frac{1}{\delta + i/2}\bm{1}
\end{equation}
We introduced the detuning parameter $\delta$ \cite{vanTiggelen2021May}, which is the difference between the frequency $\omega$ of the source and the resonance frequency $\omega_0$ of the surrounding dipoles divided by the decay rate of the dipoles. The frequency $\omega$ is assumed to be near the resonance and $\gamma \ll \omega_0$, so that we set $k=k_0$ in the Green's tensor. The T-matrix describes a non-absorbing medium, and in our numerical implementation we explicitly verified the optical theorem. The presence of an external magnetic field in the polarizability \eqref{PolarizabilityB} modifies the T-matrix \cite{vanTiggelen1996Mar}. The change is small and can be treated perturbatively. In first order, the modified T-matrix is given by
\begin{equation}
    \bm{T}_{\alpha \beta} = \bm{T}^{(0)}_{\alpha \beta} + \bm{T}^{(0)}_{\alpha \gamma }\cdot \biggl[\delta_{\gamma \delta} (- i \frac{k_0}{6 \pi} \mu \bm{\epsilon} \cdot \hat{\bm{B}}_0)\biggr] \cdot \bm{T}^{(0)}_{\delta \beta}
\end{equation}
The matrix element in polarization space $(\bm{\epsilon} \cdot \hat{\bm{B}}_0)_{ij}$ is defined as $\epsilon_{ijk}\hat{B}_{0k}$. The dimensionless parameter $\mu$ is deduced from the expression of the polarizability \eqref{PolarizabilityB} in the presence of the magnetic field:
\begin{equation}
    \mu = \frac{12 \pi}{\alpha(0) k^3_0}\frac{\omega_c}{\omega_0}
    \label{mu}
\end{equation}
All torques calculated in this paper will be linearly proportional to $\mu$. In this diamagnetic picture, $\mu \sim 5 \times 10^{-7}$ is small (calculated e.g. for hydrogen atom and the transition $1S \rightarrow 2P$).\\

The mechanical torque in Eq.~\eqref{balanceequation} can be split into two parts: the torque on the source $\bm{M}_{\text{S}}$ and the torque on the environment. The torque on the source is given by \cite{vanTiggelen2023Jan},
\begin{equation}
    M_{\text{S},i} = -\frac{2 \pi}{3} k^2 \lvert \bm{d} \rvert^2 \text{Re}\, \epsilon_{ijk}  G_{kj}^E(\bm{0},\bm{0}) 
    \label{TorqueSource}
\end{equation}
By conservation of AM, the torque on the environment equals the difference between the torque on the source and the AM leakage $  \bm{M}_{\text{E}} = \bm{M}_{\text{L}} - \bm{M}_{\text{S}}$

\section{Computation of the radiated optical angular momentum}

The right hand side of Eq.~\eqref{balanceequation} represents the radiation of optical AM as mentioned, which includes a front factor proportional to $r^3$. This factor means that the AM emitted at infinity can be found by keeping only those terms in $\bar{E}_l E_k$ and $\bar{B}_l B_k$ that are proportional to $1/r^3$ and that the leading terms $1/r^2$ vanish. We assume an unpolarised source, facilitated by a random direction of the dipole direction $\hat{\bm{d}}$ of the source. The tensor $\bar{E}_j E_m$ for an unpolarized source is expressed as:
\begin{align}
    & \bar{E}_j(\bm{r}) E_m (\bm{r})  =  \frac{(4\pi)^2}{3}k^4_0 \lvert \bm{d}\rvert^2 \biggl[G^0_{js}(\bm{r}) \bar{G}^0_{ms}(\bm{r}) \nonumber \\
    & + \sum_{\alpha,\beta} G^0_{js}(\bm{r}) \bar{G}^0_{mn}(\bm{r} - \bm{R}_\alpha) (\bar{T}_{\alpha\beta})_{nn'} \bar{G}^0_{n's}(\bm{R}_\beta) \nonumber \\
    & + \sum_{\alpha,\beta} G^0_{jn}(\bm{r} - \bm{R}_\alpha) (T_{\alpha\beta})_{nn'} G^0_{n's}(\bm{R}_\beta) \bar{G}^0_{ms}(\bm{r})  \nonumber \\
    & +\sum_{\alpha,\beta,\gamma,\delta} G^0_{jn}(\bm{r} - \bm{R}_\alpha) (T_{\alpha\beta})_{nn'} G^0_{n's}(\bm{R}_\beta) \bar{G}^0_{mp}(\bm{r} - \bm{R}_\gamma) \nonumber  \\
    & \times (\bar{T}_{\gamma \delta})_{pp'} \bar{G}^0_{p's}(\bm{R}_\delta) \biggr]
    \label{EEtensor}
\end{align}
A similar expression for average tensor $\bar{B}_j B_m$ is obtained by the same procedure. The first term of Eq.~\eqref{EEtensor} corresponds to the unscattered radiation of the source and does not contribute to the radiated AM. The second and third terms include the interference of the unscattered electric field emitted by the source and the electric field scattered by the dipole gas. The last term accounts for the interference among all multiple scattering paths inside the dipole gas. When considering the radiated optical AM at infinity, the angular integral in Eq.~\eqref{balanceequation} simplifies and can be performed analytically. The general form of the expression for the radiated AM linear in the magnetic field is given by
\begin{equation}
     M_{\text{L},k} = X_k(\bm{R}_1,...,\bm{R}_N) + A_{km}(\bm{R}_1,...,\bm{R}_N) \cdot \hat{B}_{0,m}
    \label{Leakagebeforeaveraging}
\end{equation}
The vector $\bm{X}$ vanishes when only one dipole is considered, but not for an arbitrary rotationally variant environment. The second term in Eq.~\eqref{Leakagebeforeaveraging} represents the torque linear in the magnetic field. We restore rotational invariance by averaging over the orientation of the dipole gas distribution. This means that we rotate all position vectors of the scatterers by the same angle and then calculate the average over all possible rotations. The average of $\langle \bm{M}_{\text{L}}\cdot \hat{\bm{B}}_0 \rangle_o$ over the dipole gas distribution orientation is physically equivalent to average over the direction of the magnetic field $\hat{\bm{B}}_0$, 
\begin{equation}
    \langle  \bm{M}_{\text{L}} \cdot \hat{\bm{B}}_0 \rangle_{o} = \langle  \bm{M}_{\text{L}} \cdot \hat{\bm{B}}_0 \rangle_{\hat{\bm{B}}_0}
    \label{averagingdistribution}
\end{equation}
Averaging over the orientation of the dipole gas restores rotational symmetry Consequently, only the direction of the external magnetic field remains for the radiated AM, this can be expressed as
\begin{equation}
    \langle  \bm{M}_{\text{L}}\rangle_{o} = \kappa_{\text{L}}(\bm{R}_1,...,\bm{R}_N) \hat{\bm{B}}_0
\end{equation}
The scalar $\kappa_{\text{L}}$ is easily related to the second-rank tensor $\bm{A}$ using Eqs.~\eqref{averagingdistribution} and \eqref{Leakagebeforeaveraging}:
\begin{equation}
    \kappa_{\text{L}} = \frac{1}{3}\text{Tr}\,\bm{A}
\end{equation}
The same averaging of Eq.~\eqref{TorqueSource} leads to:
\begin{equation}
    \langle \bm{M}_{\text{S}}\rangle_{o} = \kappa_{\text{S}}\hat{\bm{B}}_0
\end{equation}

The calculation of $\kappa_L$ for one single dipole, whose position is averaged over the same shell, has been done analytically. We used the analytical expression to calculate $\kappa_L$ in the single scattering approximation (see Section \ref{Numerical results}). \\

Note that $\kappa_{\text{L}}$ and $\kappa_{\text{S}}$ can still fluctuate from one realization of dipole positions $\lbrace \bm{R}_\alpha \rbrace$ to another. The full distributions will be obtained later.

\section{Orbital and Spin angular momentum}

In this section we separate the optical radiated AM into optical spin and orbital AM, and associate the total torque with the magneto-transverse component $\bm{S}_\phi$ of the Poynting vector in the far field. To establish the second we assume that the radiative AM propagates with the speed $c_0$ from the source so that,
\begin{equation}
    \frac{d \bm{J}^{\text{rad}}}{dt} = \frac{d}{dt} \biggl[\int_0^{c_0 t} dr \, r^2\int d\hat{\bm{r}} \, \bm{r} \times \frac{\bm{S}(\bm{r})}{c^2_0} \biggr]= \bm{M}_{\text{L}} 
\end{equation}
where $\bm{S}(\bm{r})$ is the Poynting vector. This equation shows that $4\pi r^3 \langle \hat{\bm{r}} \times \bm{S}_\phi \rangle_{\hat{\bm{r}}}/c_0$ for large $r$ is equal to $\bm{M}_{\text{L}}$. Hence, after averaging over the orientation of the dipole gas distribution, we get the simple expression,
\begin{equation}
    \int d\hat{\bm{r}}\, \hat{\bm{r}} \times \langle  \bm{S}_\phi \rangle_{o}(\bm{r}) = \frac{\kappa_{\text{L}}c_0}{ r^3} \hat{\bm{B}}_0
\end{equation}
We can separate the radiated AM into spin and orbital AM using \cite{Cohen-Tannoudji1987Jan},
\begin{align}
     \bm{M}_{\text{L}} & = -\frac{r^2}{8\pi k}\text{Im}\int d\hat{\bm{r}} \, \biggl[ \bm{E} \times \bar{\bm{E}} + E_m (\bm{r} \times \bm{\nabla}) \bar{E}_m \biggr] \nonumber \\
     &= \bm{M}^{\text{spin}}_{\text{L}} + \bm{M}^{\text{orbit}}_{\text{L}}
     \label{SplitSpinOrbital}
\end{align}
The angular integral in Eq.~\eqref{SplitSpinOrbital} must be evaluated on an arbitrary surface outside the environment. We evaluate it at infinity ($r\rightarrow\infty$), which allows us to perform the angular integral analytically. When evaluating the surface integral at infinity, only terms in the integrand that are proportional to $1/r^2$ need to be considered.\\

In conclusion, the scalar $\kappa_{\text{L}}$ associated with total leakage of AM, can be separated either into its spin and orbital components, or into torques exerted on environment and source,
 \begin{equation}
     \kappa_{\text{L}} = \kappa_{\text{Sp}} + \kappa_{\text{O}} = \kappa_{\text{S}} + \kappa_{\text{E}}
 \end{equation}

\section{Numerical results}
\label{Numerical results}

In the numerical simulations, we calculate the dimensionless ratios $\kappa \omega/P_{\text{S}}  \mu$. Here $P_{\text{S}}$ is the power emitted by the dipole source, which is also affected by the surrounding environment (Purcell effect \cite{Purcell}); $\kappa$ refers to either $\kappa_{\text{L}}$, $\kappa_{\text{Sp}}$, etc. This ratio normalizes the torque to the amount of radiated energy and quantifies the amount of emitted AM in units of $\hbar$ per emitted photon. The power emitted by the source is given by \cite{vanTiggelen2023Jan}:
\begin{equation}
    P_{\text{S}} = - \frac{2 \pi}{3} k_0^3 c_0\lvert \bm{d} \rvert^2 \text{Im}G_{i i}^{E}(\bm{0},\bm{0})
\end{equation}
The different ratios $ \kappa \omega/P_{\text{S}} \mu$ are computed as a function of two parameters $\eta$ and $\tau_0$ defined as,
\begin{equation}
    \eta = \frac{4 \pi n}{k_0^3} \qquad \tau_0 = \frac{R}{\ell_0} = L n \sigma_{sc}
    \label{parameters}
\end{equation}
The parameter $\eta$ is the number of dipoles per cubic of wavelength, and $\tau_0$ is (assuming independent scattering) the optical thickness, which is the ratio of the radius $R$ of the sphere and the mean free path $\ell_0 = (n\sigma_{sc})^{-1}$; $n$ is the scatterer density and $\sigma_{sc}$ is the scattering cross section for one dipole. For $\eta \ll 1$, the contribution of recurrent scattering and interference to multiple scattering is expected to be negligible, but as $\eta$ approaches unity, recurrent scattering can no longer be neglected. The neglect of recurrent scattering and interference is referred to as the Independent Scattering Approximation (ISA)\cite{DalNegro2022May}, so that $\ell_0$ is the mean free path in the independent scattering approximation. For $\eta \sim 1$, the mean free path is usually larger than $\ell_0$ \cite{vanTiggelen2021May}. We will ignore this change of the mean free path in the optical thickness $\tau_0$ and use $\ell_0$ to calculate the optical thickness. \\

 In our numerical simulations, we have excluded a small spherical region with a radius of $\lambda/4$ around the dipole source to avoid divergences due to the singular behavior of the Green's function at small $\bm{r}$ (see Fig.\ref{picture}). We accounted for this small exclusion volume to obtain the value of $\tau_0$ and $\eta$. However, we did not impose any constraints on the distance between the dipoles that would induce spatial correlations. We did not encounter any problems due to close dipoles in the simulations. To ensure the reliability of our results, we carefully ensured that the medium was on average homogeneous. We performed the full statistics of the value of the various kappa's over at least $10^3$ independent realizations for each data point, to be discussed later in Fig.\ref{Distribution}. \\

\begin{figure}[]
  \centering
  \includegraphics[width=\linewidth]{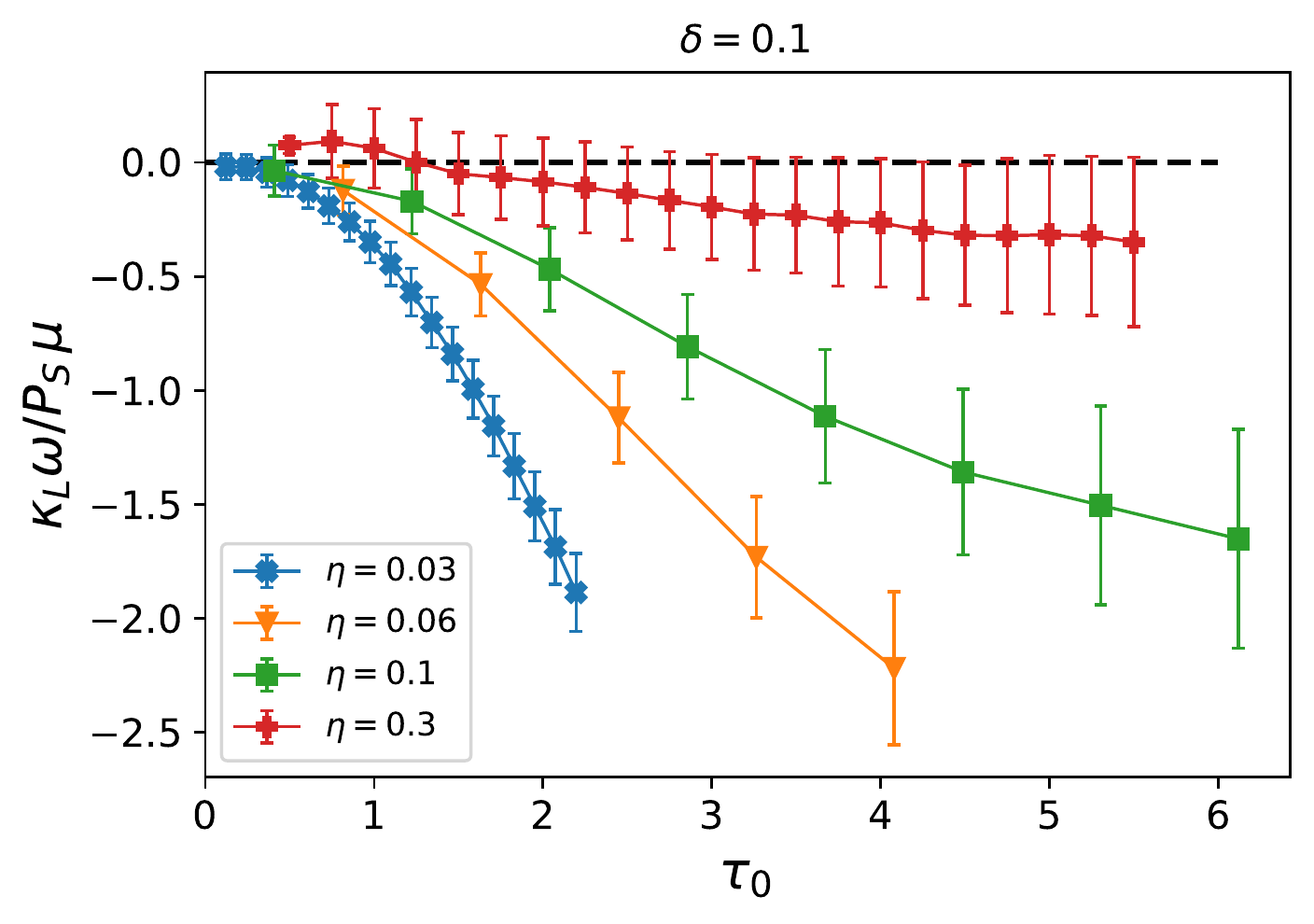}
  \caption{Numerical results for the total leakage $\kappa_{\text{L}}\omega/P_{\text{S}} \mu$ per emitted photon for different values of the parameters $\eta$ and $\tau_0$. The bars indicate the standard deviation of the full statistic of $\kappa_{\text{L}}$ values, and are not error bars. The dashed line denotes the level zero.}
  \label{KappaLeak}
\end{figure}

We now present the numerical results. Figure \ref{KappaLeak} shows that for $\eta = 0.03$ the radiated AM is proportional to $\tau^2_0$ and is directed opposite to the external magnetic field. This behavior is also expected to hold for smaller values of $\eta$ that are more difficult to access numerically. For larger $\eta$, however, the $\tau^2_0$ scaling disappears as $\tau_0$ increases, and the radiated AM decreases as $\eta$ increases at constant optical thickness. This figure illustrates the influence of recurrent scattering and interference on the radiated AM. More optical thickness increases the amount of radiated optical AM per photon, but larger values for $\eta$ reduce the amount of radiated AM. The small leak of AM for large $\eta$ may imply that interference phenomena suppress the propagation of AM inside the sphere. The result obtained for $\eta=0.3$ and $\tau_0 \sim 1.25$ is quite remarkable, since the leak of AM actually vanishes, although the torque on the source still has a significant value (Fig.\ref{KappaSourcefigure}). This indicates that a direct transfer of AM exists from the source to the environment. For these parameters, the radiated spin and orbital AM are exaclty opposite $\kappa_{\text{Sp}} + \kappa_{\text{O}} =0$, with no net leak. \\
\begin{figure}[]
  \centering
  \includegraphics[width=\linewidth]{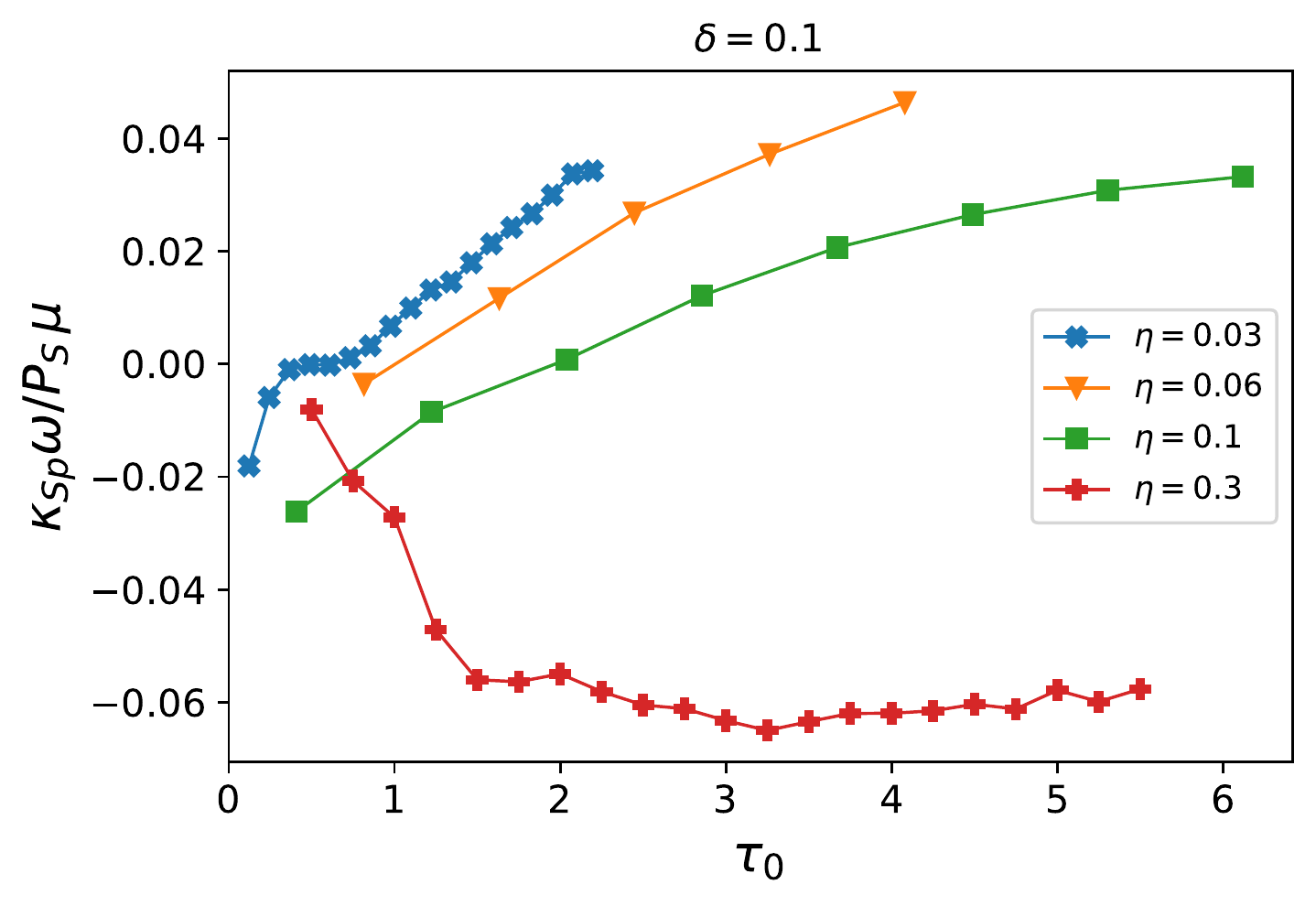}
  \caption{Results of numerical simulations for the spin leakage $\kappa_{\text{Sp}}\omega/P_{\text{S}}\mu$ per emitted photon as a function of the optical thickness for various values of the parameter $\eta$.}
  \label{kappaSpinfigure}
\end{figure}

In Fig.\ref{kappaSpinfigure} we show the numerical results for the spin component of the radiated AM. Surprisingly, we can see that the spin component represents only a small part of the total radiated AM, showing that the radiated AM is dominated by orbital AM. Only for $\eta=0.3$, the spin part of total AM becomes significant.  \\
\begin{figure}[h!]
  \centering
  \includegraphics[width=\linewidth]{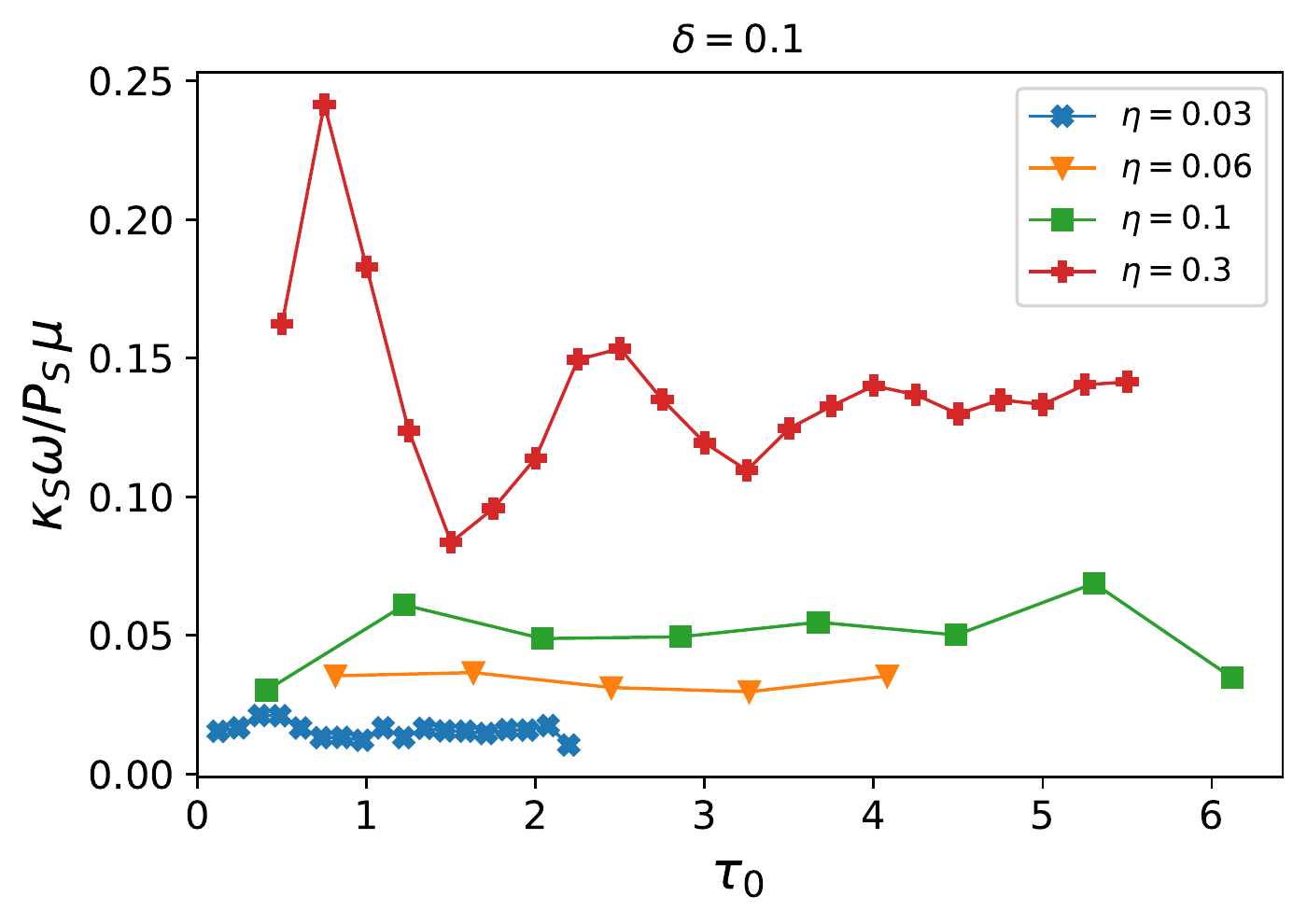}
  \caption{Numerical results showing the torque on the source per emitted photon as a function of the parameters $\eta$ and $\tau_0$.}
  \label{KappaSourcefigure}
\end{figure}

In Fig.\ref{KappaSourcefigure} we have plotted the torque acting on the source, given by the formula \eqref{TorqueSource}. It is strictly positive in the applied parameter range. Upon comparing Figs 4 and 2, it can be seen that for small values of $\eta$, the torque exerted on the source is relatively insignificant compared to the torque exerted on the environment. This implies that it is primarily the environment that acquires AM. However, as $\eta$ increases, this is no longer the case. For $\eta = 0.3$, the torque on the source becomes comparable to the torque on the environment, though with opposite sign, indicating that they acquire opposite AM. It is worth noting that the curve for $\eta=0.3$ exhibits oscillations. The cause of these oscillations is not yet fully understood, and it is possible that they are the result of an imperfect average over the dipole gas distribution.
\begin{figure}[]
  \centering
  \includegraphics[width=\linewidth]{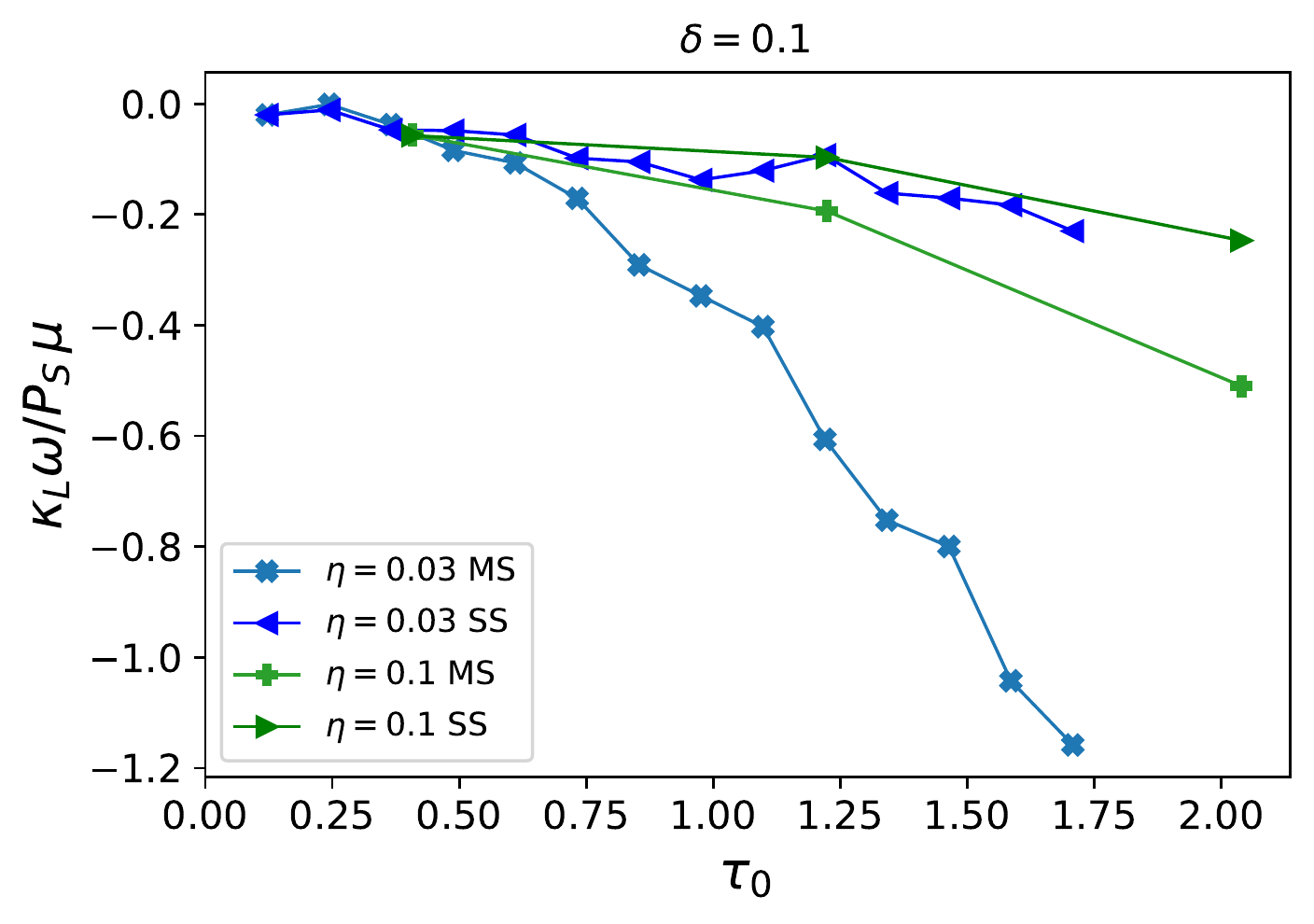}
  \caption{Effect of multiple scattering on the radiated optical AM. The multiple scattering curves are labelled 'MS' and the single scattering approximation curves are labelled 'SS'.}
  \label{Indepcompare}
\end{figure}

The aim of Fig.\ref{Indepcompare} is to compare the roles of multiple and single scattering. We compare the radiated AM calculated with multiple scattering (MS) and the radiated angular momentum calculated from $N$ single dipoles formula:
\begin{equation}
    \kappa^{\text{SS}}_{\text{L}} = \sum_\alpha \kappa_{\text{L}}(\bm{R}_\alpha)
\end{equation}
where $\kappa_{\text{L}}(\bm{R}_\alpha)$ is the radiated angular momentum for a single dipole. The single dipole formula obviously neglects any kind of multiple scattering. As expected for $\tau_0 > 1$, the multiple scattering result deviates significantly from the single dipole formula.    
\begin{figure*}[]
  \includegraphics[width=0.8\textwidth]{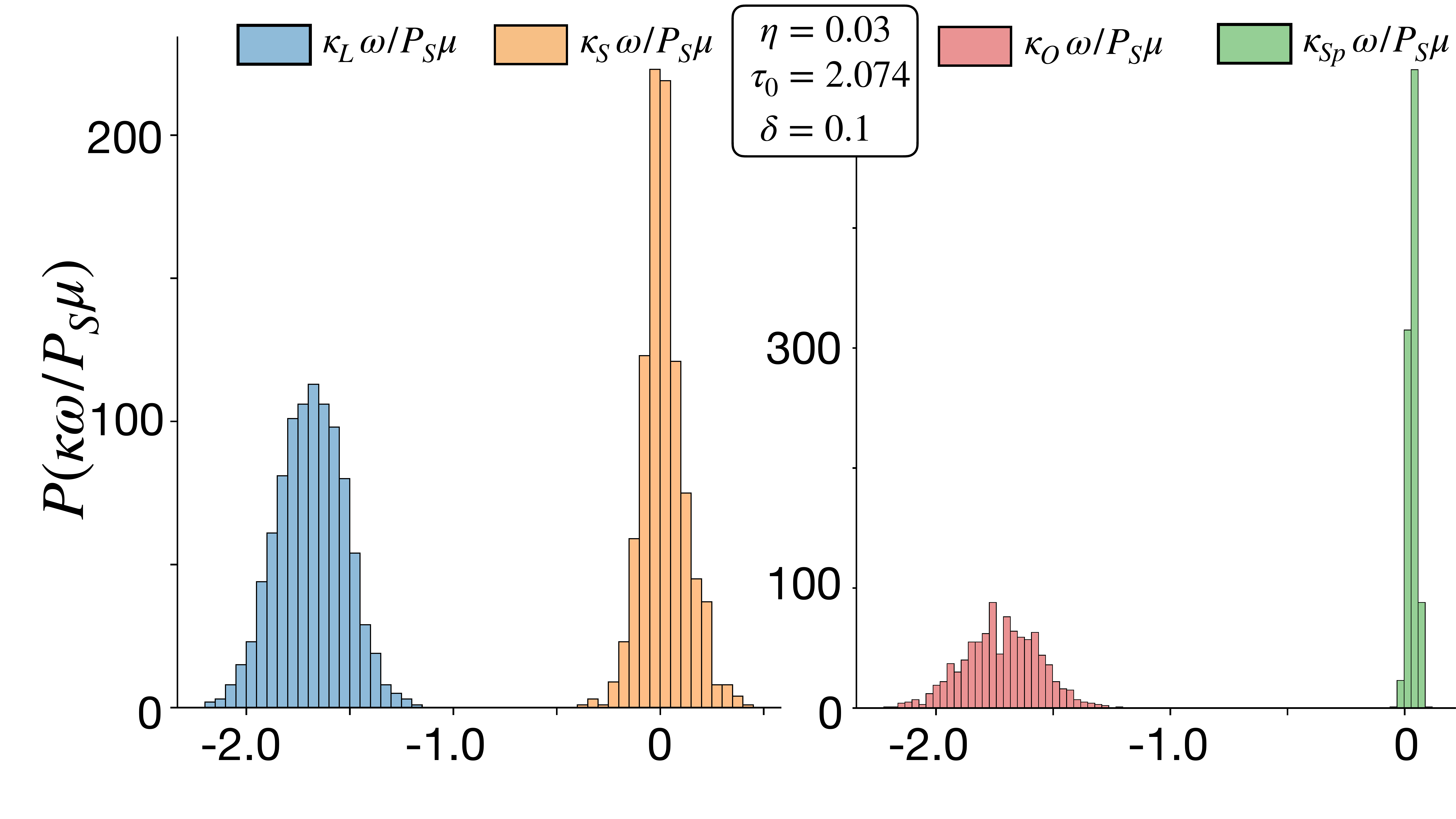}
  \caption{Left: the distribution of the normalized total leakage $\kappa_{\text{L}} \omega/P_{\text{S}} \mu$ and the normalized torque on the source $\kappa_{\text{S}} \omega/P_{\text{S}}\mu$ for $\eta = 0.03$, optical thickness $\tau_0 = 2.074$ and detuning $\delta=0.1$. The right figure depicts the distribution of the normalized spin leakage $\kappa_{\text{Sp}} \omega/P_{\text{S}} \mu$ and the normalized $\kappa_{\text{O}} \omega/P_{\text{S}} \mu$ for the same values of $\eta$, $\tau_0$ and $\delta$.}
  \label{Distribution}
\end{figure*}

The full statistics of the different $\kappa$ are shown in Fig.\ref{Distribution} for $\eta =0.03$, $\tau_0 = 2.074$ and $\delta=0.1$. The standard deviations shown earlier in Fig.\ref{KappaLeak} correspond to the widths of the probability distribution function for $\kappa_L$ for the different values of $\eta$ and $\tau_0$ and are given by the usual formula:
\begin{equation}
\sigma(\tau_0,\eta) = \sqrt{\frac{1}{N}\sum_{i=1}^N(\kappa_i - \langle \kappa \rangle)^2}
\end{equation}
where $N$ is the number of realizations, $\kappa_i$ is the value of $\kappa$ for the realization $i$ and $\langle \kappa \rangle$ is the mean value of $\kappa$ over the $N$ realizations. It can be seen from Fig.\ref{KappaLeak} that as either $\eta$ or $\tau$ increase, the width of the distributions of the radiated AM increases. Thus, both the increase in recurrent scattering and the increase in multiple scattering tend to broaden the distributions. The distributions for the spin and orbital AM are seen to be quite different, again revealing the dominant role of orbital radiated AM. Contrary to $\kappa_{\text{O}}$, the average of $\kappa_{\text{sp}}$ is small because it takes both positive and negative values. \\

\begin{figure}[]
  \centering
  \includegraphics[width=\linewidth]{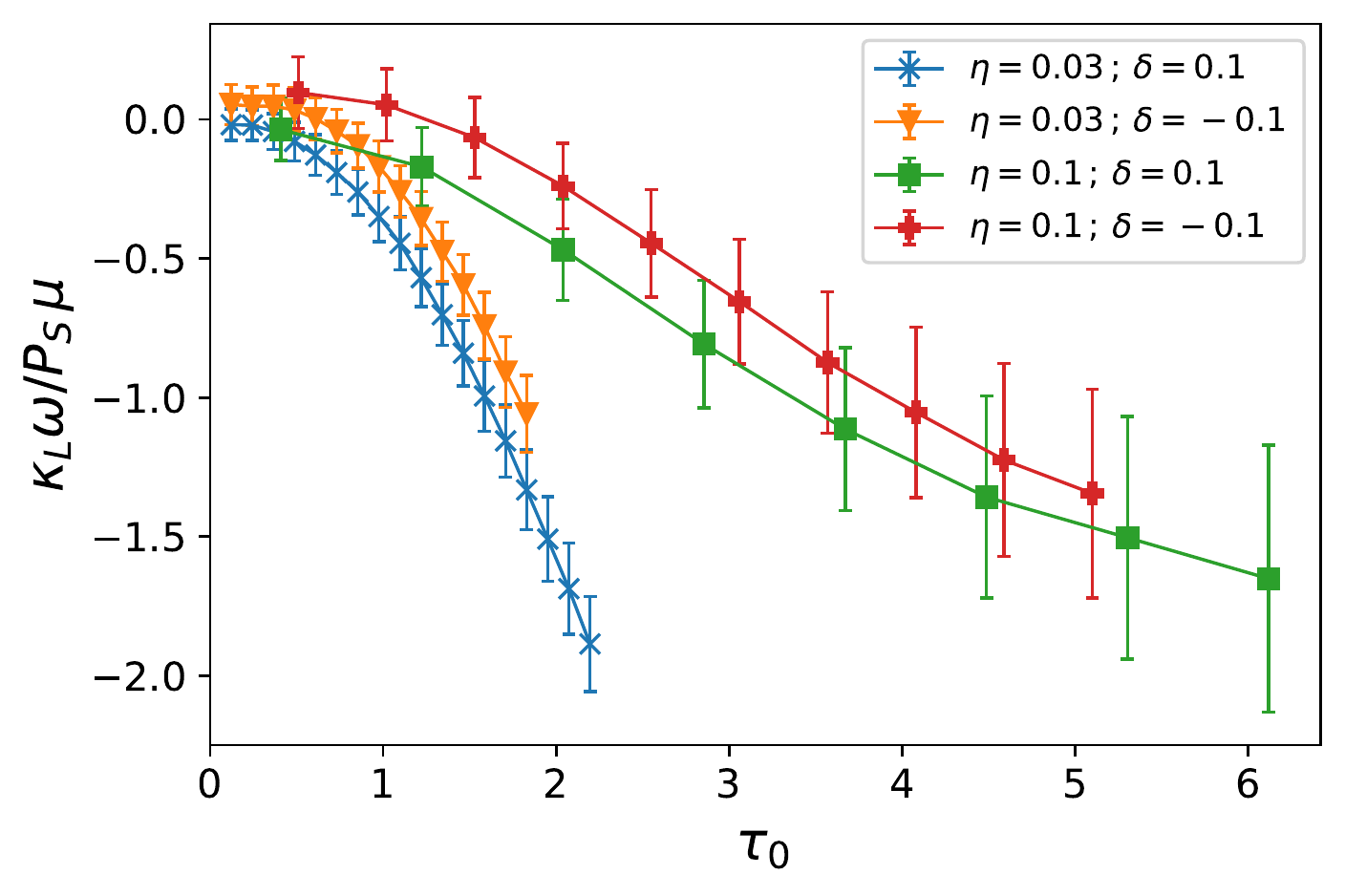}
  \caption{Impact of opposite detunings on the total AM leakage for the curves $\eta=0.03$ and $\eta=0.1$. The bars indicate the standard deviation of the distribution of $\kappa_{\text{L}} \omega/P_{\text{S}} \mu$ values.}
  \label{Oppositedetuning}
\end{figure}

In Figure \ref{Oppositedetuning}, we show the effect of taking small opposite detunings on the radiated AM for $\eta=0.03$ and $\eta=0.1$.  For small values of $\tau_0$, the radiated angular momentum has opposite signs for opposite detunings. This is expected from the single dipole formula, proportional to $\text{Im}\,t^2_0$. As $\tau_0$ approaches unity and beyond, this is no longer the case, and the curves for opposite detunings show similar behavior, with only a slight shift with respect to each other.

\begin{figure}[]
  \centering
  \includegraphics[width=\linewidth]{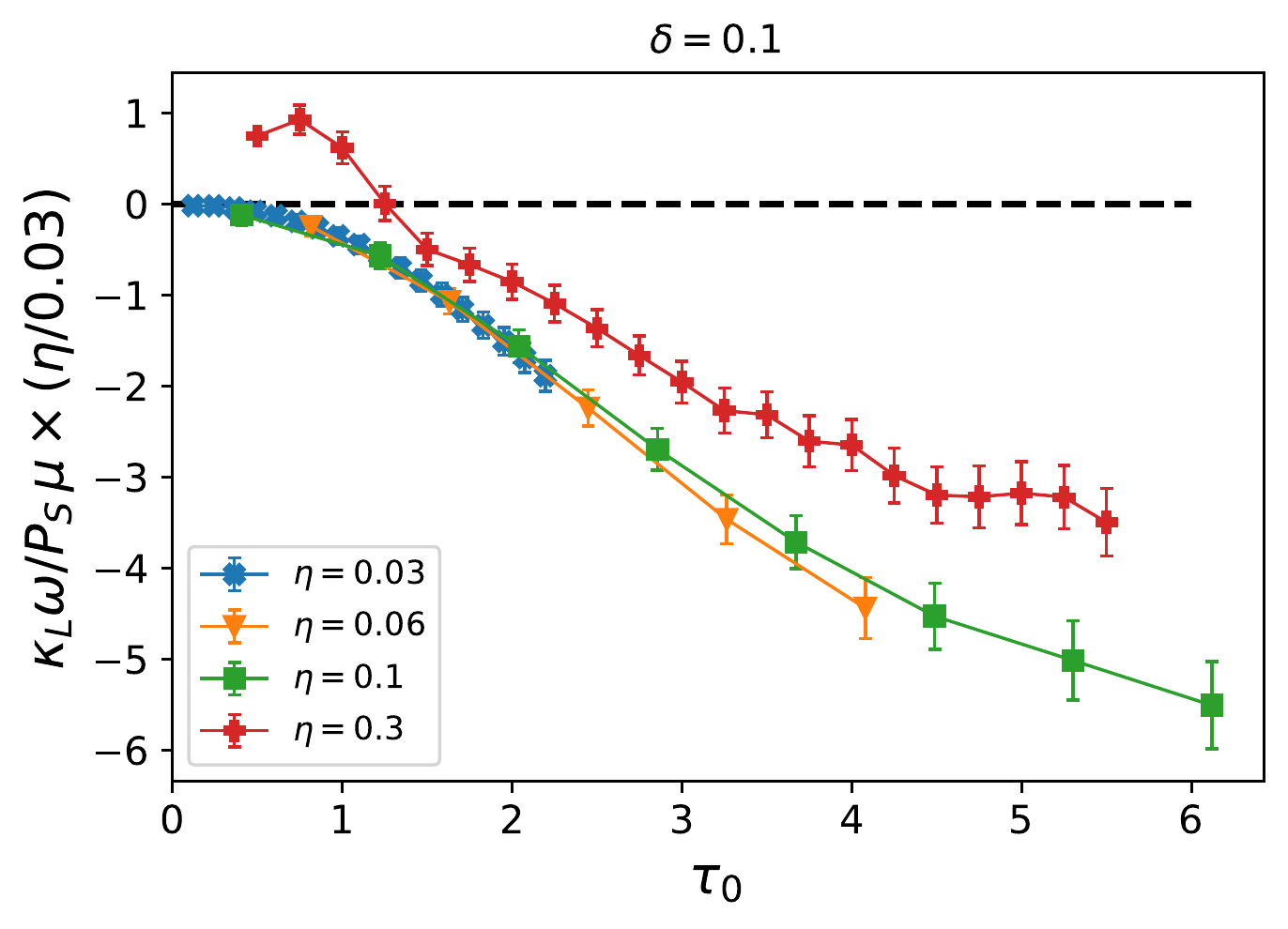}
  \caption{Single parameter scaling of the normalized angular momentum leak for $\eta = 0.03$, $\eta = 0.06$, $\eta=0.1$ and $\eta =0.3$ obtained by multiplying them by $\eta/0.03$. For $\eta = 0.3$, the scaling starts to break down. This figure shows that for values of $\eta$ up to 0.1, the total radiated angular momentum is proportional to the mean free path $\ell_0$ times $f(\tau_0)\sim \tau^2_0$ for $\tau_0 \le 2$.}
  \label{Collapse}
\end{figure}
In Fig.\ref{Collapse} we have made an attempt to find a single-parameter scaling of the AM leak. To this end we multiplied all curves by $\eta/0.03$. There is a good agreement for the curves $\eta = 0.06$ and $\eta = 0.1$ but the curve $\eta = 0.3$ clearly deviates. This figure shows that for values of $\eta$ up to 0.1, the total radiated angular momentum is proportional to the mean free path $\ell_0$ for $\tau_0$ constant. When $\eta \sim 1$, the mean free path is no more given by the ISA formula \eqref{parameters} and the true value of the optical thickness $\tau$ is smaller than $\tau_0$.

\section{Conclusion and Outlooks}

In this paper we have investigated the transfer of angular momentum radiated by an electric dipole source into a magneto-birefringent environment. Two parameters have been considered for this study: $\eta$ the density of scatterers per cubic of wavelength and $\tau_0$ the optical thickness in the independent scattering approximation (ISA). We have shown that for $\eta \ll 1$ the total radiated AM is proportional to $\tau_0^2$ and opposite to the direction of the magnetic field. However, it is observed that as $\eta$ approaches unity, the $\tau^2_0$ behavior disappears as $\tau_0$ increases. For constant $\tau_0$, the radiated AM is proportional to the mean free path $\ell_0$ as long as the density is small enough ($\eta<0.1$). \\

The torque on the medium was separated into the torque on the source and the torque on the environment. Previous work \cite{vanTiggelen2021May} showed that the torque on a homogeneous environment vanishes. For small values of $\eta$, the torque on the source is negligible and the torque environment dominates. As $\eta$ increases, the torque on the source increases while the total radiated AM decreases. We conclude that there is more and more transfer of AM from the source directly to the environment. For $\tau_0 \sim 1.25$ and $\eta =0.3$ we see a special case with no radiated AM at all, thus the torque on the source is fully compensated by the torque on the environment. Finally, the impact of small opposite detunings $\delta$ on the total radiated AM is studied.  \\

We have also separated the total radiated angular momentum into spin and orbital components. Surprisingly, the radiated angular momentum turns out to be dominated in most cases by orbital AM. \\

This work can be extended to consider a magnetic dipole field rather than a uniform magnetic field, without much qualitative difference. This could be relevant for astrophysical phenomena such as AM transport in stars.

\section*{Acknowledgements}

This work was funded by the Agence Nationale de la Recherche (Grant No. ANR-20-CE30-0003 LOLITOP). \\

The structure of this text has been improved with the help of DeepL write.


\nocite{Skipetrov2014Jan,Rusek1997Oct,Leseur2014Nov,Skipetrov2018Aug}

\newpage
\bibliographystyle{apsrev}
\bibliography{Projetdipole}

\begin{thebibliography}{14}
\expandafter\ifx\csname natexlab\endcsname\relax\def\natexlab#1{#1}\fi
\expandafter\ifx\csname bibnamefont\endcsname\relax
  \def\bibnamefont#1{#1}\fi
\expandafter\ifx\csname bibfnamefont\endcsname\relax
  \def\bibfnamefont#1{#1}\fi
\expandafter\ifx\csname citenamefont\endcsname\relax
  \def\citenamefont#1{#1}\fi
\expandafter\ifx\csname url\endcsname\relax
  \def\url#1{\texttt{#1}}\fi
\expandafter\ifx\csname urlprefix\endcsname\relax\def\urlprefix{URL }\fi
\providecommand{\bibinfo}[2]{#2}
\providecommand{\eprint}[2][]{\url{#2}}

\bibitem[{\citenamefont{Purcell}(1946)}]{Purcell}
\bibinfo{author}{\bibfnamefont{E.~M.} \bibnamefont{Purcell}},
  \bibinfo{journal}{Phys. Rev.} \textbf{\bibinfo{volume}{69}},
  \bibinfo{pages}{681} (\bibinfo{year}{1946}).

\bibitem[{\citenamefont{van Tiggelen and Rikken}(2020)}]{vanTiggelen2020Sep}
\bibinfo{author}{\bibfnamefont{B.~A.} \bibnamefont{van Tiggelen}}
  \bibnamefont{and} \bibinfo{author}{\bibfnamefont{G.~L. J.~A.}
  \bibnamefont{Rikken}}, \bibinfo{journal}{Phys. Rev. Lett.}
  \textbf{\bibinfo{volume}{125}}, \bibinfo{pages}{133901}
  (\bibinfo{year}{2020}).

\bibitem[{\citenamefont{van Tiggelen}(2023)}]{vanTiggelen2023Jan}
\bibinfo{author}{\bibfnamefont{B.~A.} \bibnamefont{van Tiggelen}},
  \bibinfo{journal}{Opt. Lett.} \textbf{\bibinfo{volume}{48}},
  \bibinfo{pages}{41} (\bibinfo{year}{2023}).

\bibitem[{\citenamefont{Barron}(2004)}]{Barron2004Sep}
\bibinfo{author}{\bibfnamefont{L.~D.} \bibnamefont{Barron}},
  \emph{\bibinfo{title}{{Molecular Light Scattering and Optical Activity}}}
  (\bibinfo{publisher}{Cambridge University Press},
  \bibinfo{address}{Cambridge, England, UK}, \bibinfo{year}{2004}).

\bibitem[{\citenamefont{van Tiggelen and Skipetrov}(2021)}]{vanTiggelen2021May}
\bibinfo{author}{\bibfnamefont{B.~A.} \bibnamefont{van Tiggelen}}
  \bibnamefont{and} \bibinfo{author}{\bibfnamefont{S.~E.}
  \bibnamefont{Skipetrov}}, \bibinfo{journal}{Phys. Rev. B}
  \textbf{\bibinfo{volume}{103}}, \bibinfo{pages}{174204}
  (\bibinfo{year}{2021}).

\bibitem[{\citenamefont{Landau and Lifshitz}(1980)}]{Landau1980Jan}
\bibinfo{author}{\bibfnamefont{L.~D.} \bibnamefont{Landau}} \bibnamefont{and}
  \bibinfo{author}{\bibfnamefont{E.~M.} \bibnamefont{Lifshitz}},
  \emph{\bibinfo{title}{{The Classical Theory of Fields: Volume 2}}}
  (\bibinfo{publisher}{Butterworth-Heinemann}, \bibinfo{address}{Oxford,
  England, UK}, \bibinfo{year}{1980}).

\bibitem[{\citenamefont{Newton}(2013)}]{Newton2013Jun}
\bibinfo{author}{\bibfnamefont{R.~G.} \bibnamefont{Newton}},
  \emph{\bibinfo{title}{{Scattering Theory of Waves and Particles: Second
  Edition (Dover Books on Physics)}}} (\bibinfo{publisher}{Dover Publications},
  \bibinfo{address}{Mineola, NY, USA}, \bibinfo{year}{2013}).

\bibitem[{\citenamefont{van Tiggelen et~al.}(1996)\citenamefont{van Tiggelen,
  Maynard, and Nieuwenhuizen}}]{vanTiggelen1996Mar}
\bibinfo{author}{\bibfnamefont{B.~A.} \bibnamefont{van Tiggelen}},
  \bibinfo{author}{\bibfnamefont{R.}~\bibnamefont{Maynard}}, \bibnamefont{and}
  \bibinfo{author}{\bibfnamefont{T.~M.} \bibnamefont{Nieuwenhuizen}},
  \bibinfo{journal}{Phys. Rev. E} \textbf{\bibinfo{volume}{53}},
  \bibinfo{pages}{2881} (\bibinfo{year}{1996}).

\bibitem[{\citenamefont{Cohen-Tannoudji
  et~al.}(1987)\citenamefont{Cohen-Tannoudji, Dupont-Roc, and
  Grynberg}}]{Cohen-Tannoudji1987Jan}
\bibinfo{author}{\bibfnamefont{C.}~\bibnamefont{Cohen-Tannoudji}},
  \bibinfo{author}{\bibfnamefont{J.}~\bibnamefont{Dupont-Roc}},
  \bibnamefont{and} \bibinfo{author}{\bibfnamefont{G.}~\bibnamefont{Grynberg}},
  \emph{\bibinfo{title}{{Photons et atomes - Introduction
  {\ifmmode\grave{a}\else\`{a}\fi}
  l'{\ifmmode\acute{e}\else\'{e}\fi}lectrodynamique quantique}}}
  (\bibinfo{publisher}{EDP Sciences}, \bibinfo{year}{1987}).

\bibitem[{\citenamefont{Dal~Negro}(2022)}]{DalNegro2022May}
\bibinfo{author}{\bibfnamefont{L.}~\bibnamefont{Dal~Negro}},
  \emph{\bibinfo{title}{{Waves in Complex Media}}}
  (\bibinfo{publisher}{Cambridge University Press},
  \bibinfo{address}{Cambridge, England, UK}, \bibinfo{year}{2022}).

\bibitem[{\citenamefont{Skipetrov and Sokolov}(2014)}]{Skipetrov2014Jan}
\bibinfo{author}{\bibfnamefont{S.~E.} \bibnamefont{Skipetrov}}
  \bibnamefont{and} \bibinfo{author}{\bibfnamefont{I.~M.}
  \bibnamefont{Sokolov}}, \bibinfo{journal}{Phys. Rev. Lett.}
  \textbf{\bibinfo{volume}{112}}, \bibinfo{pages}{023905}
  (\bibinfo{year}{2014}).

\bibitem[{\citenamefont{Rusek et~al.}(1997)\citenamefont{Rusek, Or{\l}owski,
  and Mostowski}}]{Rusek1997Oct}
\bibinfo{author}{\bibfnamefont{M.}~\bibnamefont{Rusek}},
  \bibinfo{author}{\bibfnamefont{A.}~\bibnamefont{Or{\l}owski}},
  \bibnamefont{and}
  \bibinfo{author}{\bibfnamefont{J.}~\bibnamefont{Mostowski}},
  \bibinfo{journal}{Phys. Rev. E} \textbf{\bibinfo{volume}{56}},
  \bibinfo{pages}{4892} (\bibinfo{year}{1997}).

\bibitem[{\citenamefont{Leseur et~al.}(2014)\citenamefont{Leseur, Pierrat,
  S{\ifmmode\acute{a}\else\'{a}\fi}enz, and Carminati}}]{Leseur2014Nov}
\bibinfo{author}{\bibfnamefont{O.}~\bibnamefont{Leseur}},
  \bibinfo{author}{\bibfnamefont{R.}~\bibnamefont{Pierrat}},
  \bibinfo{author}{\bibfnamefont{J.~J.}
  \bibnamefont{S{\ifmmode\acute{a}\else\'{a}\fi}enz}}, \bibnamefont{and}
  \bibinfo{author}{\bibfnamefont{R.}~\bibnamefont{Carminati}},
  \bibinfo{journal}{Phys. Rev. A} \textbf{\bibinfo{volume}{90}},
  \bibinfo{pages}{053827} (\bibinfo{year}{2014}).

\bibitem[{\citenamefont{Skipetrov and Sokolov}(2018)}]{Skipetrov2018Aug}
\bibinfo{author}{\bibfnamefont{S.~E.} \bibnamefont{Skipetrov}}
  \bibnamefont{and} \bibinfo{author}{\bibfnamefont{I.~M.}
  \bibnamefont{Sokolov}}, \bibinfo{journal}{Phys. Rev. B}
  \textbf{\bibinfo{volume}{98}}, \bibinfo{pages}{064207}
  (\bibinfo{year}{2018}).

\end{thebibliography}

\end{document}